\documentclass[10pt,conference,letterpaper]{IEEEtran}
\usepackage{etex}

\usepackage{etoolbox}
\makeatletter
\patchcmd{\maketitle}{\@copyrightspace}{}{}{}
\makeatother

\usepackage[utf8]{inputenc}

\usepackage{mathptmx}
\usepackage[english]{babel}
\usepackage{multirow}
\usepackage{rotating}
\usepackage{color}
\usepackage{xspace}
\usepackage{tabularx}
\usepackage{balance}
\usepackage{tikz}
\usepackage{comment}
\usepackage{pgfplots}
\usepgfplotslibrary{units}
\usepackage{caption}
\usepackage{framed}
\usepackage{mdframed}
\usepackage{latexsym}
\usepackage{amssymb}
\usepackage{amsmath}
\usepackage{subcaption} 
\usepackage[inline]{enumitem}
\usepackage{xcolor}
\usepackage{algorithm}
\usepackage{algorithmic}

\usepackage{listings} 
\usepackage{textcomp}
\usepackage[bookmarks=false]{hyperref}
\usepackage{cite}

\usetikzlibrary{arrows}

\definecolor{primary}{RGB}{117,112,179}
\definecolor{secondary}{RGB}{27,158,119}
\definecolor{tertiary}{RGB}{217,95,2}

\pgfplotsset{compat=newest
           , tick label style = {font=\tiny}
}

\definecolor{orange}{RGB}{255,127,0}
\definecolor{grey}{RGB}{135,135,135}

\let\origthelstnumber\thelstnumber

\lstdefinelanguage{diff}{
  language=java,
  basicstyle=\ttfamily\scriptsize,
  sensitive=true,
  morecomment=[f][\color{gray}][0]{diff},
  morecomment=[f][\color{gray}][0]{index},
  morecomment=[f][\color{blue}][0]{@@},
  morecomment=[f][\color{magenta}][0]{***},
  morecomment=[f][\color{violet}][0]{!},
  morecomment=[f][\color{red!60!black}][0]{-},
  morecomment=[f][\color{green!60!black}][0]{+},
  morecomment=[f][\color{magenta}][0]{---},
  morecomment=[f][\color{magenta}][0]{+++},
  morecomment=[f][\color{gray}][0]{Binary},
  morecomment=[f][\color{gray}][0]{Only},
  morecomment=[f][\color{gray}][0]{old},
  morecomment=[f][\color{gray}][0]{new},
  morecomment=[f][\color{gray}][0]{rename},
  morecomment=[f][\color{gray}][0]{similarity},
  morecomment=[f][\color{gray}][0]{deleted},
  morecomment=[f][\color{magenta}][0]{***************},
  morecomment=[f][\color{red!60!black}][0]<,
  morecomment=[f][\color{green!60!black}][0]>,
  morecomment=[f][\color{blue}][0]{0},
  morecomment=[f][\color{blue}][0]{1},
  morecomment=[f][\color{blue}][0]{2},
  morecomment=[f][\color{blue}][0]{3},
  morecomment=[f][\color{blue}][0]{4},
  morecomment=[f][\color{blue}][0]{5},
  morecomment=[f][\color{blue}][0]{6},
  morecomment=[f][\color{blue}][0]{7},
  morecomment=[f][\color{blue}][0]{8},
  morecomment=[f][\color{blue}][0]{9},
}[comments]
\makeatletter
\newcommand*\Suppressnumber{
  \lst@AddToHook{OnNewLine}{
    \let\thelstnumber\relax
     \advance\c@lstnumber-\@ne\relax
    }
}

\newcommand*\Reactivatenumber[1]{
  \setcounter{lstnumber}{\numexpr#1-1\relax}
  \lst@AddToHook{OnNewLine}{
   \let\thelstnumber\origthelstnumber
   \refstepcounter{lstnumber}
  }
}
\makeatother

\AtBeginDocument{

}

\usepackage[]{pdfcomment}

\newcommand{\question}[2]{{RQ#1. #2}}
\newcommand{\answer}[2]{\vspace{.3cm}{\centering\fbox{\parbox{0.95\columnwidth}{\textbf{Answer to RQ#1}. #2}}}\vspace{.3cm}}
\newcommand{\definition}[2]{{Definition: \bf #1} #2}
\newcommand{\tabincell}[2]{\begin{tabular}{@{}#1@{}}#2\end{tabular}}

\newcommand\algName{FO-EXPLORE\xspace}
\newcommand{\mycode}[1]{{\small \texttt{#1}}\xspace}

\newcommand\nbBugs{16\xspace}

\newcommand\stratReplaceVar{S1\xspace}

\newcommand\stratReplaceNew{S2\xspace}

\newcommand\stratSkipLine{S3\xspace}
\newcommand\stratReturnNull{S4a\xspace}
\newcommand\stratReturnVar{S4c\xspace}
\newcommand\stratReturnNew{S4b\xspace}

\newcommand\nbValidRuntimePatches{8460\xspace}

\newcommand\thetitle{Exhaustive Exploration of the Failure-oblivious Computing Search Space}

\newcommand\rqCore{\question{1}{[Multiplicity] Does it exist multiple failure-oblivious decision sequences for a given failure-triggering input? How large is the corresponding search space?}\xspace}
\newcommand\rqProportionValid{\question{2}{[Fertility] What is the proportion of \underline{valid} failure-oblivious decision sequences?}\xspace}
\newcommand\rqSizeDisparity{\question{3}{[Disparity] To what extent does the search space contain composite decision sequences?}\xspace}

\begin{document}
\title{\thetitle}

\author{\IEEEauthorblockN{Thomas Durieux\IEEEauthorrefmark{1}, 
        Youssef Hamadi\IEEEauthorrefmark{2}, 
        Zhongxing Yu\IEEEauthorrefmark{1}, 
        Benoit Baudry\IEEEauthorrefmark{3},
        Martin Monperrus\IEEEauthorrefmark{3}} 
        \IEEEauthorblockA{
        \IEEEauthorrefmark{1}University of Lille \& Inria, France 
        \IEEEauthorrefmark{2}Ecole Polytechnique, France}
        \IEEEauthorrefmark{3}Royal Institute of Technology, Stockholm}

\maketitle

\begin{abstract}
High-availability of software systems requires automated handling of crashes in presence of errors.
Failure-oblivious computing is one technique that aims to achieve high availability.
We note that failure-obliviousness has not been studied in depth yet, and there is very few study that helps understand why failure-oblivious techniques work.
In order to make failure-oblivious computing to have an impact in practice, we need to deeply understand failure-oblivious behaviors in software.
In this paper, we study, design and perform an experiment that analyzes the size and the diversity of the failure-oblivious behaviors.
Our experiment consists of exhaustively computing the search space of 16 field failures of large-scale open-source Java software.
The outcome of this experiment is a much better understanding of what really happens when failure-oblivious computing is used, and this opens new promising research directions.
\end{abstract}

\section{Introduction}

Dependable computing  is founded on four main engineering activities \cite{avizienis2004basic}:
fault prevention;
fault removal, i.e. testing;
fault forecasting;
and fault tolerance.
The latter, fault tolerance, aims at handling faults triggered in production. 
For instance, it is common to observe error pages on the Internet while ordering a laser pointer, registering to a conference, or installing a new blogging platform.
Fault tolerance aims at preventing system crashes on occurrence of such errors. As such, it is complementary to testing, as it takes care of the faults that have not been discovered at testing time.

Failure-oblivious computing is one of the approaches for fault tolerance \cite{rinard2004enhancing,dobolyi2008changing,long2014automatic,berger2006diehard}.
To overcome software failures at runtime, failure-oblivious computing techniques modify the execution so that failures become less critical: instead of crashing the whole system, only the current task fails and the system remains available. 
For example, Rinard et al. \cite{rinard2004enhancing} have proposed a failure-oblivious model consisting of skipping erroneous writes happening out of an array's bounds. Another example is probabilistic memory safety in DieHard \cite{berger2006diehard}, where the execution modifications are controlled addition of blank memory padding. 

In this work, we address the following main limiting assumption of existing works on failure-oblivious computing : 
previous work in this area assumes that in response to a given failure, there exists one and only one execution modification to be applied.
In practice, however, there can possibly exist multiple failure-oblivious decisions to respond to a failure. As an example, consider the null pointer exceptions in a high-level language such as Java.
To overcome the same null pointer exception, one can imagine two different solutions: 1) one can skip the execution of the statement where a null value is about to be used, or 2) one can directly return to the caller method. 
We scientifically investigate this assumption in this paper. More specifically, we set up a conceptual failure-oblivious computing framework in which this assumption does not hold, i.e., several execution modifications exist within this framework to respond to a failure.

Once we accept that there are multiple alternative execution modifications to the same failure, this naturally leads to the novel and promising concept of ``search space of failure-oblivious computing''. The basic search space for a single failure is simply composed of all possible execution modifications. 
However, now consider the presence of multiple failures occurring in a row in a single execution, then the search space becomes the cartesian product of all possible decisions over each failure point. To exhaustively explore this search space, we propose an algorithm called \algName.

We then setup an experiment to systematically study the failure-oblivious search space of \nbBugs field failures of large-scale open-source Java software. 
We run those field failures in a virtual endless loop that simulates the same failure happening again and again.
Each execution offers us a possibility to explore a new patch in the search space, where a patch is composed by a sequence of execution modifications. The outcome of the experiment is the exact topology, for a given failure-oblivious model, of the failure-oblivious
computing search space of the considered failures.

The experimental result shows that the failure-oblivious search space is in general large and not easy to explore: this experiment represents more than 10 days of computation in a distributed scientific grid.
For all the \nbBugs considered field failures, we identify a total of \nbValidRuntimePatches different failure-oblivious execution modifications which are composed of 1 up to 8 execution modifications taken in a row. 
As an example, let us consider a field failure of Apache Commons Collections, an open-source Java project.
This field failure, reported as issue \#360, is a null pointer exception.
For this failure, our proposed exploration algorithm \algName finds that there exist 45 different failure-oblivious execution modifications.

To sum up, the contributions of this paper are:
\begin{itemize}

  \item The characterization of the concept of ``search space of failure-oblivious computing''. 
  
  \item An algorithm to exhaustively explore this novel search space.

  \item An implementation of the algorithm for exhaustively exploring the failure-oblivious computing search space for null pointer dereferences in Java.

  \item The systematic empirical study of the failure-oblivious computing search space for \nbBugs real null dereference failures, reported on a public issue tracker on large and used Java libraries.
\end{itemize}

The remainder of this paper is organized as follows.
\autoref{sec:motivation} motivates the study of the failure-oblivious computing search space.
\autoref{sec:contribution} details our exploration algorithm.
\autoref{sec:evaluation} details the evaluation on \nbBugs field null dereferences.
\autoref{sec:threats_validity} details the threats to validity of the contribution.
\autoref{sec:rw} presents the related works and Section \ref{sec:conclusion} concludes.
This paper is a complete rewrite of an Arxiv version \cite{DBLP:journals/corr/DurieuxHM16}.

\section{Motivating Example}
\label{sec:motivation}

Let us consider the example in \autoref{fig:npe-example}, which is an excerpt of server code that retrieves the last connection date of a user and prints the result to an HTML page.
Method \mycode{getLastConnectionDate} first gets the user session, and then retrieves the last connection date from the session object.
This code snippet can possibly trigger two failures that can crash the request:
1) if the session does not exist and \mycode{getUserSession} returns null, then there is a null pointer exception at line 3 (NPE1), and
2) for the first connection, \mycode{getLastConnection} returns null, and another null pointer exception can be thrown at line 6 (NPE2).

Now let us consider the possible execution modifications to overcome the failures.
In \autoref{fig:npe-example}, to overcome NPE1 at line 3, a failure-oblivious system could modify the execution state and flow in three ways:
1) it creates a new session object on the fly, and
2) it returns an arbitrary Date object such as the current date, and
3) it returns null.
As the example suggests, \emph{there are multiple possible failure-oblivious strategies for the same failure}.
However, note that not all such modifications are equivalent. For instance, if modification \#3 is applied, it triggers another failure NPE2, whereas solutions \#1 and \#2 do not further break the system state. This indicates that not all state modifications are equivalent, some being invalid.

In this paper, we define a conceptual framework and an exploration algorithm to reason about the presence of multiple competing execution modifications in response to a failure.

\begin{lstlisting}[numbers=left, caption={Code Excerpt with Two Potential Null Dereference Failures},label=fig:npe-example,float=b,floatplacement=b]
Date getLastConnectionDate() {
    Session session = getUserSession();
    return session.getLastConnection(); // NPE1
}
...
HTML.write(getLastConnectionDate().toString()); // NPE2
\end{lstlisting}

\section{Exploring the Search Space of Failure-oblivious Computing}
\label{sec:contribution}

We have shown that there is an implicit search space in  failure-oblivious computing. In this section, we formally define this search space and devise an algorithm to explore it exhaustively.

\subsection{Basic Definitions}

Failure-oblivious computing \cite{rinard2004enhancing} is the art of changing a program's state or flow during execution such that a crashing failure does not happen anymore and that the program is able to continue its execution, and it is sometimes referred to as runtime repair \cite{Lewis2010}, state repair \cite{Monperrus2015} and self-healing software \cite{Locasto2006}. As the terminology of failure-oblivious computing can hardly be considered as stable, we thus first define its core terms and concepts.

\definition{A failure point}{is the location in the code where a failure is triggered. In the simplest case, a failure point is a statement at a given line.}

\definition{A failure-oblivious model}{defines a type of failure and the corresponding possible manners to overcome the failure. For instance, a failure-oblivious model that considers out-of-bounds write in arrays can skip the array upon failures.
}

\definition{A failure-oblivious strategy}{defines how a failure is handled. The traditional failure-oblivious literature assumes that there is one single way to be oblivious to a failure. However, there are types of failures for which there exist multiple failure-oblivious strategies. For example, upon an out-of-bounds read in an integer array, one can return either a constant or a value that presents somewhere else in the array. This makes two different strategies.}

\definition{A context-dependent failure-oblivious strategy}{is a strategy that can be instantiated in multiple ways depending on the execution context. For example, upon an out-of-bounds read in an integer array, one can return the first, the second, \ldots, up to the $n^{th}$ value that presents in the array, meaning that there are \textit{n} ways to instantiate this strategy.} 

\definition{A failure-oblivious decision}{is the application of a specific failure-oblivious strategy to handle a failure at a specific failure point. A failure-oblivious decision is an execution modification.}

\subsection{The Failure-oblivious Computing Search Space}
    
Taking a failure-oblivious decision means exploring a new program state after the first failure. The execution proceeding from this new program state can result in a new failure, for which a failure-oblivious decision has to be taken as well.

\definition{A failure-oblivious decision sequence}{is a sequence of failure-oblivious decisions that are taken in a row during one single execution because of cascading failures.}

In this paper, an execution consisting of a  single failure-oblivious decision is called a unary decision sequence, and a composite (or n-ary) decision sequence contains at least 2 decisions.
We use \autoref{fig:npe-example} to give an example of n-ary failure-oblivious decision sequence. At line 3, the execution of the method can be stopped by returning null to overcome a null pointer exception when session is null, later on in the execution, a new Date can be crafted to handle the second null pointer exception at line 6. 
In certain cases, only one decision in isolation may not be enough to overcome the failure, only the sequence is a solution. 

The notion of context-aware decision and decision sequence naturally defines a search space.

\definition{The failure-oblivious computing search space}{ of a failure-triggering input is defined by all possible decision sequences that can happen after the first failure.}

\subsection{\algName: An Algorithm to Explore the Failure-oblivious Computing Search Space}

After defining the failure-oblivious computing search space, we now present an algorithm to exhaustively explore this search space. The algorithm is named \algName and is shown in \autoref{algo:top-level}.

\subsubsection{Algorithm Input-Output}
\label{sec:algorithm}

\algName requires four inputs which we explain in detail below.

\textit{Program P}: a program to which failure-oblivious support will be injected. The program is automatically transformed so as to support failure-oblivious execution.
The transformation also adds intercession hooks to steer and monitor failure-oblivious decisions.

\textit{Failure Triggering Input I}: an input triggering a runtime failure. We assume that we have at least one program input that enables us to automatically reproduce the failures as many times as we want. An input can be a set of values given to a function. In an object-oriented program, an input is much more than values only, it is a set of created objects, interacting through a sequence of method calls.

\textit{Failure-oblivious Model R}: a model listing the possible modifications on the program state or execution flow to handle the failure as defined above.

\textit{Validity Oracle O}: an oracle specifying the viability of the computation if failure-oblivious computing happens.
A validity oracle is a predicate on the program state at the end of program execution on the failure-triggering input.
The goal of the validity oracle is to validate or invalidate the failure-oblivious decision sequence that has happened during execution.
In failure-oblivious computing, the traditional validity oracle is the absence of crashing errors, but more advanced ones can be defined (e.g., use additional assertions).
For instance, consider again the example in \autoref{fig:npe-example} where the validity oracle is the presence of an exception that results in HTTP code 500.
Returning null is a failed unary decision sequence because the request crashes with NPE1. 
On the contrary, returning a fresh date object enables the request to succeed and the HTML to be generated, making it a valid unary decision sequence.

Taking the above described four inputs, \algName outputs a set of failure-oblivious decision sequences along with their validity.

\subsubsection{Algorithm Workflow}
\begin{algorithm}[t]
  \caption{The core exploration protocol \algName}
  \label{algo:top-level}

  \begin{algorithmic}[1]
    
    \REQUIRE{P: a program}
    \REQUIRE{I: an input for P}
    \REQUIRE{R: a failure-oblivious model}
    \REQUIRE{O: a validity oracle}
    \ENSURE{S: a set of failure-oblivious decision sequences}
    \STATE{$failurePoints \leftarrow \emptyset$}
    \STATE{$decisions \leftarrow map_{failurePoint \rightarrow decision}$}

    \WHILE{last decision sequence was unknown}
        \STATE{$seq \leftarrow \emptyset$}
        \STATE{exec P(I)}
        
        \WHILE{failure $f$ happens according to R}
           \IF{$f \notin failurePoints$}
                \STATE{$failurePoints \leftarrow failurePoints \cup f$}
                \STATE{$decisions[f]  \leftarrow computePossibleDecisions()$}
            \ENDIF
                \STATE{select an unexplored or a not completely explored decision $d$}
                \STATE{$seq \leftarrow seq + d$}
                
                \STATE{apply d (change the state or flow)}
            \STATE{proceed with execution}
        \ENDWHILE
        \STATE{store pair $(decision sequences, O(seq))$ in S}
    \ENDWHILE
  \end{algorithmic}
\end{algorithm}

To explore the search space, the basic idea behind \algName is to make a different decision according to the failure-oblivious model under consideration for each time a failure point is detected, and then collect all decision sequences.

We now give a detailed description of the workflow. To explore all the failure-obvious decisions sequences (line 3), \algName executes the input that triggers the runtime failure as much as it is required (line 5).
Each time that the failure is detected at a failure point ($f$) according to the failure-oblivious model (line 6),
\algName checks whether it has already chosen a decision for the failure point ($f$) (line 7).
If no decision has been taken for the current failure point ($f$) before, \algName computes all the possible decisions for the current failure point (line 9)\footnote{\algName uses the original NPEFix \cite{durieux2017dynamic} algorithm to compute the search space of a single failure point.}. 
Then \algName selects an unexplored or a not completely explored decision ($d$) from the computed available decisions (line 11).
More specifically, \algName can face two different selection cases:\\
\textbf{Case 1:} The failure point has never been detected before, which means that the failure has never happened in this location before. 
In this case, \algName selects the first decision in the set of possible $decisions$.\\
\textbf{Case 2:} The failure has already been detected before at this failure point in the program. 
In this case, \algName has two possibilities: explore a new failure-oblivious decision or use an already used decision that triggers a new failure point which is not exhaustively explored.

Once a decision has been selected, \algName will store it in the current decision sequence (line 12).
Then \algName applies the decision and resumes the execution of the program (lines 13-14).
At the end of the execution, \algName uses the validity oracle to determine the validity of the execution and finally stores the result (line 16).

\begin{figure}
\centering
\includegraphics[width=0.97\columnwidth]{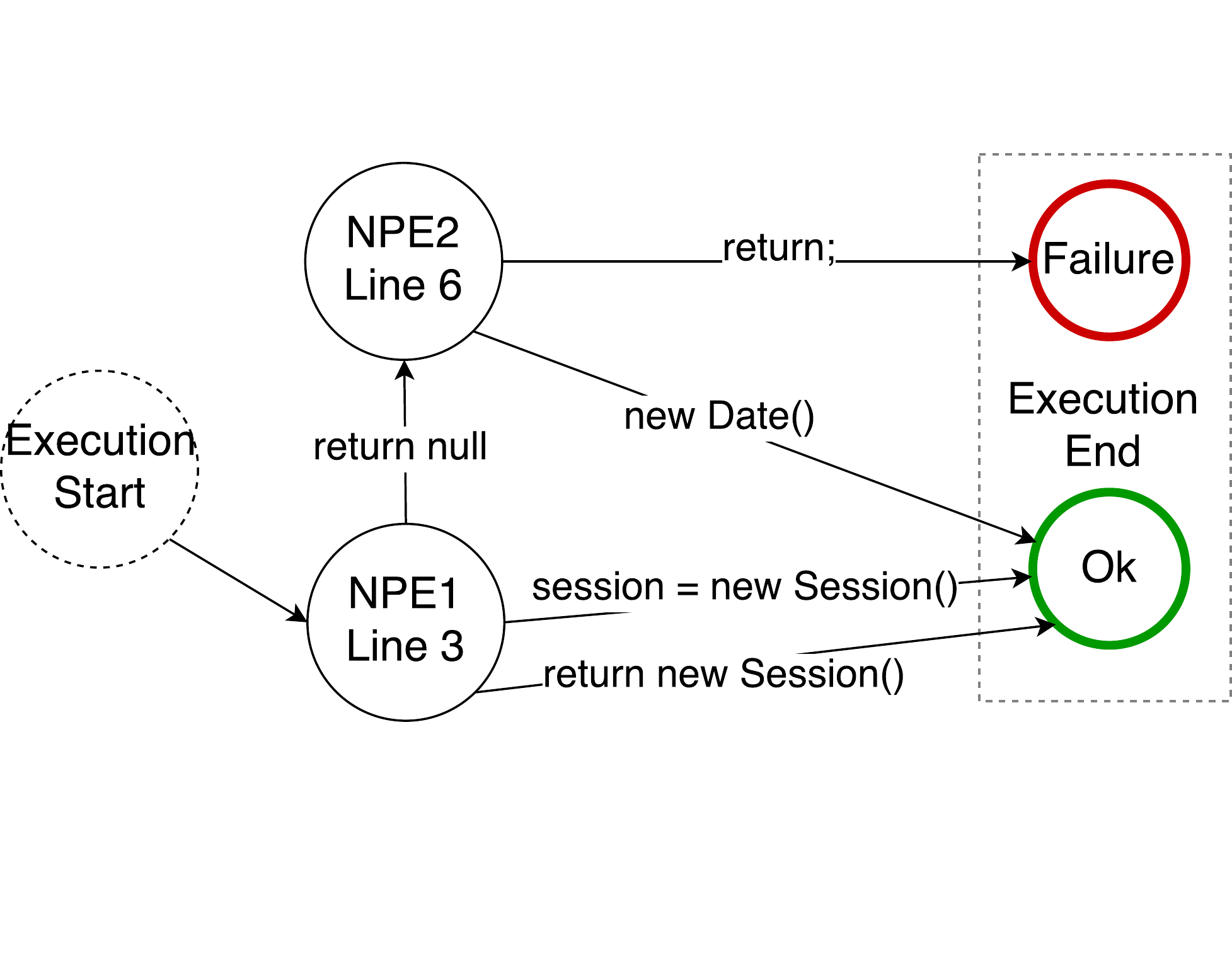}
\caption{Excerpt of the decision tree of the example \autoref{fig:npe-example}. One path is this tree is a ``decision sequence'''. NPEX refers to a failure point.}
\label{fig:decisionTreeExample} 
\end{figure}

\subsubsection{Working Example}
We now illustrate the workflow of \algName using the example in \autoref{fig:npe-example}. \autoref{fig:decisionTreeExample} presents the actual decision tree of this example.
Recall that when \algName executes the request for the failing input, a null pointer exception called NPE1 will be produced at line 3. This is shown as a circle ``NPE1'' in \autoref{fig:decisionTreeExample}.
For this first failure point (NPE1), \algName explores three different decisions to handle the null pointer exception (the three arrows coming out of NPE1 in \autoref{fig:decisionTreeExample}): 1) uses \mycode{return null} to exit the \mycode{getLastConnectionDate} method, 2) uses \mycode{return new Date()} to exit the method, and 3) uses \mycode{session = new Session()} to initialize the null variable with a new instance and proceeds with the execution of the same method.
In cases 2) and 3), the execution does not produce other exceptions.

Now consider that \algName selects the first decision (\mycode{return null}) and resumes the execution.
This decision later produces a second null pointer exception (NPE2 in \autoref{fig:decisionTreeExample}) at line 6 in \autoref{fig:npe-example}.
At this failure point, \algName explores two decisions: 1) uses \mycode{return} to exit the execution of the method and 2) uses \mycode{new Date()} to replace the null expression \mycode{getLastConnectionDate()} with a new instance of \mycode{Date}.
The latter execution modification succeeds while the former produces a failure. The former execution modification fails as if \algName selects the \mycode{return} strategy at failure point of NPE2, 
no response is sent to the client so that a timeout will be produced on the client side.

At the end of the exploration, \algName eventually discovers four different decision sequences (the four different paths from Execution Start to Execution End in \autoref{fig:decisionTreeExample}), and three of them produce an acceptable output.

\subsection{Usefulness of Exploring the Search Space}
\label{sec:benefits}

Characterizing the search space of failure-oblivious computing can provide sound scientific foundations on future work on failure-oblivious computing. Indeed, there is very little work that studies to what extent and why failure-oblivious computing succeeds. By clearly defining and exploring the search space, we obtain comprehensive data about this unresearched phenomenon. The empirical results presented in \autoref{sec:evaluation} is a first step towards this direction.

In addition, by identifying multiple acceptable failure-oblivious decision sequences, it opens a radically new perspective on failure-oblivious computing: there may be certain decision sequences that are better than the others. In other words, there are cases where one adds an additional criterion on top of the validity oracle, and this criterion is used to select the best failure-oblivious decision sequence. For instance, the best decision sequence may be the one that runs the fastest. In presence of multiple acceptable decision sequences,  one needs to characterize and explore the search space first in order to select the best failure-oblivious decision sequence.

\section{Empirical Evaluation} \label{sec:evaluation}

The goal of the empirical evaluation presented in this section is to study the failure-oblivious computing search space of real failures occurring in large-scale open-source software. 
This study is built on three research questions about the topology of the search space.

\rqCore
We want to understand if it is possible to apply different failure-oblivious strategies to handle one specific failure. And if it is the case, how many different decision sequences can be taken.

\rqProportionValid
In the context of failure-oblivious computing, there may exist different failure-oblivious decision sequences that are all valid, i.e., they all fix the runtime failure under consideration. 
To us, the proportion of valid decision sequences can be considered as the ``fertility'' of the search space.
When the goal is to find at least one valid decision sequence, it is much easier to do so if there are many points in the search space that are valid.
On the contrary, if there is only one valid decision sequence in the search space, it requires in the worst case visiting the complete search space before finding the only valid decision sequence.
The fertility of the search space is the opposite of what is called ``hardness'' or ``constrainedness'' in combinatorial optimization.

\rqSizeDisparity
Our protocol identifies a set of diverse failure-oblivious decision sequences, and some of them require several decisions in a row.
We will observe whether there exist such composite failure-oblivious decision sequences in our benchmark.

\subsection{Considered Failure-Oblivious Model}
\label{sec:implementation}

\begin{table}
\centering
\caption{In our experiments, we consider a failure-oblivious model for null dereference exceptions in Java. The model consists of 6 possible strategies.}
\label{tab:strategies}
\resizebox{0.97\columnwidth}{!}{
\begin{tabularx}{\columnwidth}{|l|l|l|l|X|}
\hline
\multicolumn{3}{|c|}{Strategy} &  Id & Description \\ \hline
\multicolumn{2}{|c|}{\multirow{3}{*}{replacement}}
 & reuse               & \stratReplaceVar   & injection of an existing compatible object \\ \cline{3-5}
\multicolumn{2}{|c|}{} & {creation}         & \stratReplaceNew  & injection of a new object \\ \hline
\multirow{6}{*}{\rotatebox{90}{skipping}} 
 & \multicolumn{2}{c|}{line}                & \stratSkipLine    & skip statement \\ \cline{2-5}
 & \multirow{4}{*}{\rotatebox{90}{method}} 
                                 & void     & \stratReturnNull  & return a null or $\emptyset$ to caller \\ \cline{3-5}
 &                               & creation & \stratReturnNew   & return a new object to caller \\ \cline{3-5}
 &                               & reuse    & \stratReturnVar   & return an existing compatible object to caller \\
\hline
\end{tabularx}
}
\end{table}

We have implemented \algName for null dereferences in Java (aka null pointer exceptions) with a failure-oblivious model derived from NPEFix \cite{durieux2017dynamic}.
In Java, all object variable dereferences are potential failure points (field accesses, method calls on local variables, method parameters, and implicit casts).
A decision has to be taken when the variable is null, which means that failure points are activated if and only if a null is going to be dereferenced.
For each decision point, NPEFix defines six failure-oblivious strategies, grouped into two categories as shown in \autoref{tab:strategies}. 
The first category consists of replacing the null value by an alternative valid non-null object of a compatible type.
This category is composed of two-sub-categories:
1) when a variable is null, one can reuse an object from another variable in the scope instead, strategies of this kind are called reuse-based strategies;
2) when a variable is null, one can also create a new object on the fly, strategies of this kind are called creation-based strategies. 
Note that the number of possible decisions for reuse and creation based strategies is parametrized by the number of variables and the number of constructors available, which means that for a single decision point, there are often dozens of different available decisions.
The second category is based on skipping the execution of the code affected by the null dereference, one can either skip the line that uses the null dereference or skip the rest of the method. When skipping the rest of a method which returns a value, one can also either reuse an existing object or create one on the fly. 

For more details on the strategies and the implementation, we refer the readers to the NPEFix paper \cite{durieux2017dynamic}.

\subsection{Benchmark}\label{sec:dataset}
\begin{table}[t]
\centering
\caption{Dataset of \nbBugs bugs with null dereference in six Apache open-source projects.}
\label{tab:dataset}
\resizebox{0.98\columnwidth}{!}{
\setlength\tabcolsep{0.7 ex}
\begin{tabular}{|l|l|r|r|}\hline
Bug ID          & SVN revision& LOC   & \tabincell{c}{\# method calls \\ before null}\\\hline
Collections-360 & 1076034     & 21650 &                         13 \\
Felix-4960      & 1691137     & 33057 &                          2 \\
Lang-304        & 489749      & 17277 &                          2 \\
Lang-587        & 907102      & 17317 &                         10 \\
Lang-703        & 1142381     & 19047 &                          9 \\
Math-1115       & 1590254     & 90782 &                        328 \\
Math-1117       & 1590251     & 90794 &                        342 \\
Math-290        & 807923      & 38265 &                         88 \\
Math-305        & 885027      & 38893 &                          8 \\
Math-369        & 940565      & 41082 &                          7 \\
Math-988A       & 1488866     & 82442 &                        136 \\
Math-988B       & 1488866     & 82443 &                        134 \\
PDFBox-2812     & 1681643     & 67294 &                         37 \\
PDFBox-2965     & 1701905     & 64375 &                         54 \\
PDFBox-2995     & 1705415     & 64821 &                         37 \\
Sling-4982      & 1700424     & 1182  &                          2 \\
\hline
\multicolumn{2}{|l|}{\tabincell{l}{Total: \nbBugs failures from \\6 open-source large projects}} & 770721  & 1209\\
\hline
\end{tabular}
}
\end{table}

We need real and reproducible production failures to conduct the evaluation. 
To achieve this, we reuse the benchmark of NPEFix \cite{durieux2017dynamic} consisting of \nbBugs field failures coming from Apache projects.
\autoref{tab:dataset} presents the core statistics of our benchmark.
The first column contains the bug id.
The second column contains the SVN revision of the global Apache SVN.
The third column contains the number of lines of code.
The fourth column contains the total number of method calls before the null pointer exception is triggered.
For example, issue Collections-360 fixed at revision 1076034 is an application of 21650 lines of Java code.

Let us dwell on the last column, the number of executed method calls before the dereference happens.
It gives an intuition on the complexity of the setup required to reproduce the field failure. As shown in \autoref{tab:dataset}, there are between 2 and 342 methods (application methods, not counting JDK and API methods) called for the reproduced field failures under consideration, with an average of 75.56. This indicates that the failures in our benchmark are not simple tests with a trivial call to a method with null parameters.  

\subsection{Experimental Protocol}
\label{sec:protocol}

The experiment is based on the exhaustive exploration of the search space of failure-oblivious decision sequences, as defined by our failure-oblivious model for null pointer exceptions described in \autoref{sec:implementation}.
It is done on the benchmark of failures presented in \autoref{sec:dataset}.

\subsubsection{Exhaustive exploration}
\label{sec:exhaustive-exploration}

We apply algorithm \algName{} to build the complete decision tree of all failures in our benchmark. 
Recall that the exploration of the failure-oblivious research space is done as follows:
\begin{enumerate*}
  \item We instrument each buggy program of our benchmark with our failure-oblivious model;
  \item We execute each instrumented program with the test case that encodes the field failure;
  \item We collect all decisions taken at runtime; and finally
  \item We execute the validity oracle at the end of the test case execution.
\end{enumerate*}

The time required to perform such an experiment is approximately the size of the search space multiplied by the time for reproducing the failure.
Note the alternative computation that comes after the first failure-oblivious decision at the first failure point is added on top of this.

The raw data of this evaluation is publicly available on GitHub\footnote{ \url{https://github.com/Spirals-Team/runtime-repair-experiments/}}.
We answer all research questions based on this data.

\subsubsection{Validity of Failure-Obliviating Decision Sequences}
\label{sec:validity}
For a given decision sequence taken in response to a failure, we assess its validity according to the oracle. 
In our experiments, the validity oracle is directly extracted from the test case reproducing the field failure:
a decision sequence is considered as valid if no null pointer exception and no other exceptions are thrown.
A decision sequence is considered as invalid if the original null pointer exception is thrown (meaning that there is no possible decision at the failure point), or another exception is thrown and not caught.
When the test case contains domain-specific assertions beyond whether exceptions occur, we keep them and a decision sequence is considered as valid if all assertions pass after the application execution modifications. This is the case for 14/\nbBugs failures.

\subsection{Results to Research Questions}
\subsubsection{\rqCore}
\label{sec:rq1}

\begin{table*}[t]
\centering
\caption{Key metrics of the failure oblivious search space according to \algName.}
\label{tab:space}
{
\setlength\tabcolsep{1.5 ex}
\begin{tabularx}{0.8\textwidth}{|X|r|r|r|r|r|r|r|r|r|r|}\hline
                            & \multicolumn{2}{c|}{RQ1}  & \multicolumn{5}{c|}{RQ2/RQ3}  \\\cline{2-8} 
\multirow{6}{*}{Bug ID}     & \multirow{6}{*}{\rotatebox{0}{\tabincell{c}{Nb encountered \\ decision points}}}
                            & \multirow{6}{*}{\rotatebox{0}{\tabincell{c}{Nb  possible \\ decision \\ sequence}}} 
                            & \multirow{6}{*}{\rotatebox{0}{\tabincell{c}{Nb valid \\ decision \\ sequence}}}
                            & \multirow{6}{*}{\rotatebox{0}{\tabincell{c}{\% decision \\ sequence}}}
                            & \multicolumn{3}{c|}{\multirow{5}{*}{\tabincell{c}{$\mid$ Valid decision seq..$\mid$}}} \\
                            & & & & & \multicolumn{3}{l|}{} \\
                            & & & & & \multicolumn{3}{l|}{} \\
                            & & & & & \multicolumn{3}{l|}{} \\
                            & & & & & \multicolumn{3}{l|}{} \\\cline{6-8}
                            & & & & &  Min. & Med. & Max.  \\
                            \hline
Collections-360 &      2 &     45 &     16 & 35,5\% &      2 &      2 &      2  \\
Felix-4960      &      1 &     10 &      4 &  40\% &      1 &      1 &      1  \\
Lang-304        &      1 &      7 &      6 & 35,5\% &      1 &      1 &      1  \\
Lang-587        &      1 &     28 &      1 & 3,0\% &      1 &      1 &      1  \\
Lang-703        &      4 &    459 &    130 & 28,3\% &      2 &      2 &      2  \\
Math-1115       &      1 &      5 &      5 &  100\% &      1 &      1 &      1  \\
Math-1117       &     21 &  51785 &   7708 & 14,9\% &      7 &      8 &      8  \\
Math-290        &      1 &     14 &      4 & 28,6\% &      1 &      1 &      1  \\
Math-305        &      1 &      4 &      3 &   75\% &      1 &      1 &      1  \\
Math-369        &      2 &     14 &      0 &    0\% &  ---   &  ---   &  ---    \\
Math-988A       &      3 &    576 &    383 & 66,5\% &      1 &      2 &      3  \\
Math-988B       &      1 &     32 &     17 & 53,1\% &      1 &      1 &      1  \\
Pdfbox-2812     &      8 &    294 &    168 & 57,1\% &      1 &      6 &      7  \\
Pdfbox-2965     &      1 &      4 &      3 &   75\% &      1 &      1 &      1  \\
Pdfbox-2995     &      1 &      5 &      1 &   20\% &      1 &      1 &      1  \\
Sling-4982      &      2 &     16 &     11 & 68,7\% &      1 &      1 &      1  \\
\hline
Total           &     51 &  53298 &   8460 & 15,9\% &      1 &      1 &      8  \\
\hline
\end{tabularx}
}
\end{table*}

We analyze the data obtained with the experiment described in \autoref{sec:exhaustive-exploration}, consisting of exhaustively exploring the search space of execution modifications for \nbBugs null dereferences.
\autoref{tab:space} shows the core metrics we are interested in.  

\autoref{tab:space} reads as follows.
Each line corresponds to a failure of our benchmark. Each column gives the value of a metric of interest. 
The first column contains the name of each bug.
The second column contains the number of encountered decision points for this failure.
The third column contains the number of possible failure-oblivious decision sequences for this failure.
The fourth column contains the number of execution modifications (valid decision sequences for which the oracle has stated that the decision sequence has worked).
The fifth column contains the percentage of valid decision sequences.
the sixth column contains the minimum/median/maximum number of decisions taken for valid decision sequences.

For example, the first line of \autoref{tab:space} details the result for bug Collections-360.
To overcome this failure at runtime, there are two possible decision points, which, when they are systematically unfolded, correspond to 45 possible decision sequences, 16 of which are valid according to the oracle.
The size of the valid decision sequences is always equal to $2$, which means that there must be two decisions taken in a row to handle the failure. 

Our experiment is the first to show that there exist multiple alternative decisions to overcome a failure at runtime, as shown by the number of explored decision sequences. The number is exactly the size of our search space when we conduct an exhaustive exploration. 
In this experiment, it ranges between 4 decisions (for Math-305 and PdfBox-2965) to 576 for Math-988A and 51785 for Math-1117.
Overall, we notice a great variance of the size of the search space.

We also see in \autoref{tab:space} that there is a correlation between the number of activated decision points for a given failure and the number of possible decision sequences. 
For instance, for Felix-4960, there is only one activated decision point (at the failure point where the null pointer exception is about to happen), and 10 possible decisions can be taken at this point. 
On the contrary, Math-1117 has the biggest number of activated decision points (21), which also has the biggest number of decision sequences (51785).
This correlation is expected and explained analytically as follows.
Once a first decision is made at the failure point (where the null dereference is about to happen), many alternative execution paths are uncovered. Then, a combinatorial explosion of stacked decisions happens. If we assume that there are 5 alternative decisions at the first decision point and that each of them in turn triggers another decision point with 10 alternative decisions, it directly results in $5\times 10=50$ possible decision sequences. 
Now, if we assume $n$ decision points with $m$ possible decisions on average, this results in $m^n$ decision sequences, which is a combinatorial explosion.
In general, the size of the search space depends on:
\begin{enumerate*}
\item the overall number of decision points activated for a given failure,
\item the number of possible decisions at each decision point,
\item and the correlation between different decision points, that is the extent to which one decision taken at a decision point influences the number of possible subsequent decisions to be taken.
\end{enumerate*}
For failures with large number of explored decision sequences, it means that failure-oblivious computing unfolds a large number of diverse program states and their corresponding subsequent executions.

\answer{1}{We have performed an exhaustive exploration to draw a precise picture of the failure-oblivious computing search space. 
The result clearly shows that there exist multiple alternative failure-oblivious decision sequences to handle null dereferences. 
In our experiment, there are 11/\nbBugs failures for which we observe more than 10 possible decision sequences (column ``Nb possible decision sequence'') for the same failure and according to our execution modification model, with a maximum value of 51785 (for Math-1117).}

\subsubsection{\rqProportionValid}
\label{sec:rq2}
We now have a clear picture of the size of the failure-oblivious search space, and are interested in further knowing whether there exist multiple valid decision sequences in that space.
To do so, we still consider the exhaustive study protocol described in \autoref{sec:exhaustive-exploration} whose results are given in \autoref{tab:space}.
We concentrate especially on the fourth column which shows the number of valid decision sequences. We compare it against the column representing the size of the search space, i.e., the total number of possible decision sequences. 
For instance, for Collections-360, the search space contains 45 possible decision sequences, among which 16 are valid according to the oracle (the absence of null pointer exception and the two assertions at the end of the test case reproducing the failure pass), i.e., a proportion of $16/45=36\%$ of decision sequences in the search space are valid.

We notice several interesting extreme cases in \autoref{tab:space}.
First, there are two failures -- Lang-587 and PdfBox-2995 -- for which only 1 valid decision sequence exists.
In addition, there is one failure for which all decision sequences remove the failure. This is Math-1115 for which all 5 possible decision sequences are valid.

Let us dwell on this proportion of valid decision sequences in the search space. This proportion depends on the strength of the considered oracle.
In failure-oblivious computing, the oracle that is classically considered is the absence of runtime exceptions: we call this oracle the default oracle. 
In this experiment, we have an oracle that subsumes this default oracle as we also use assertions at the end of the test that reproduces the failure.
We have manually inspected the tests and found that not all tests have equally strong assertions, which partly explains the variations in fertility we observe. 

\answer{2}{In our benchmark, the proportion of valid decision sequences varies from 0 to 100\% (from 0/14 to 5/5 valid execution modifications). This great variation is due to strength of the considered oracle, and the complexity of the code at the failure point.}

\subsubsection{\rqSizeDisparity}
\label{sec:rq3}

We have shown in {\bf RQ2} that there are multiple valid execution modifications.
Now, we study their complexity as measured by the number of decisions involved in the execution modification.
To do so, we again study the results of the exhaustive study protocol described in \autoref{sec:exhaustive-exploration} whose results are given in \autoref{tab:space}.
We especially concentrate on the column showing that among the set of valid decision sequences, what is the minimum, median and maximum size (recall that the size is the number of decisions in the decision sequence).
For instance, for PDFBox-2812, the minimal size in number of decisions among all valid failure-oblivious decision sequences is $1$, the median size is $6$ and the maximal size is $7$.

We have the following findings from this data.
First, for $5/\nbBugs$ failures of our benchmark, we see that there exist failure-oblivious decision sequences composed of more than one decision. Since our failure-oblivious model is specific to null pointer exceptions, it means that there exist failures for which the null dereference problem is not solved by the first decision, and that another null dereference happens later. Among the 5 failures, the failure-oblivious decision sequences are of the same size 2 for two failures (Collections-360, Lang-703). For the remaining 3 failures (Math-988A, PDFBox-2812, Math-1117), there are failure-oblivious decision sequences of different size. For instance, for Math-988A, there exist failure-oblivious decision sequences of 1, 2 and 3 decisions.
Second, for $10/\nbBugs$ failures of our benchmark, the failure-oblivious decision sequences are always composed of a single decision. This is strongly correlated with the size of the search space (third column, number of decision sequences), indicating that the test case reproducing the production failure sets up a program state that is amenable to failure-oblivious computing.
Finally, for one failure (Math-369), even though there are several decisions taken, but none of them are valid. All decision sequences are invalidated by the oracle (the assertions of the test case reproducing the field failure).

\begin{figure}
\centering
\begin{tikzpicture}
  \begin{axis}[
    ybar=0pt,
    grid=both,
    grid style={line width=.1pt, draw=gray!10},
    y axis line style = { opacity = 0 },
    xlabel=\# Strategies involved,
    ylabel=\# Valid decision sequences,
    tickwidth         = 1pt,
    bar width=0.75cm,
    nodes near coords,
  ]
  \addplot[fill=grey, area legend] coordinates { (1,171) (2,321) (3,166) (4,63) (5,2)};
  \end{axis}
\end{tikzpicture}
\caption{The number of valid decision sequences that use between 1 and 5 different repair strategies. We exclude Math-1117 for better visibility.}
\label{fig:nb_strat_nb_decision}
\end{figure}
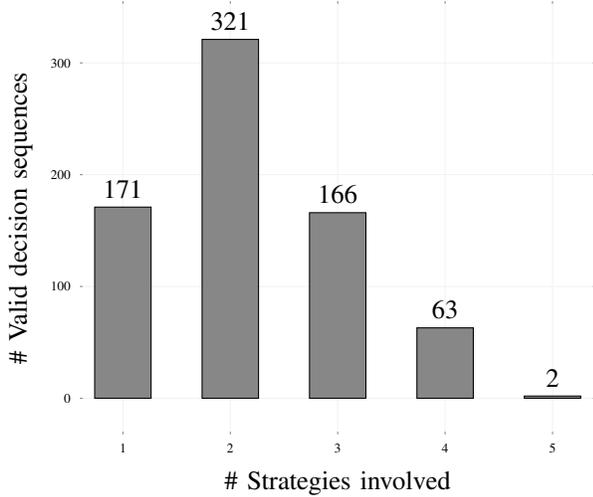

An interesting question is whether one needs to apply different strategies in the same execution to get an acceptable result.
\autoref{fig:nb_strat_nb_decision} presents the number of different repair strategies that are used in valid decision sequences.
We see that 76\% of valid decision sequences mix at least two different strategies, this clearly hints that mixing different strategies is important for effective failure oblivious computing. 

\begin{figure}
\centering
\includegraphics[width=0.97\columnwidth]{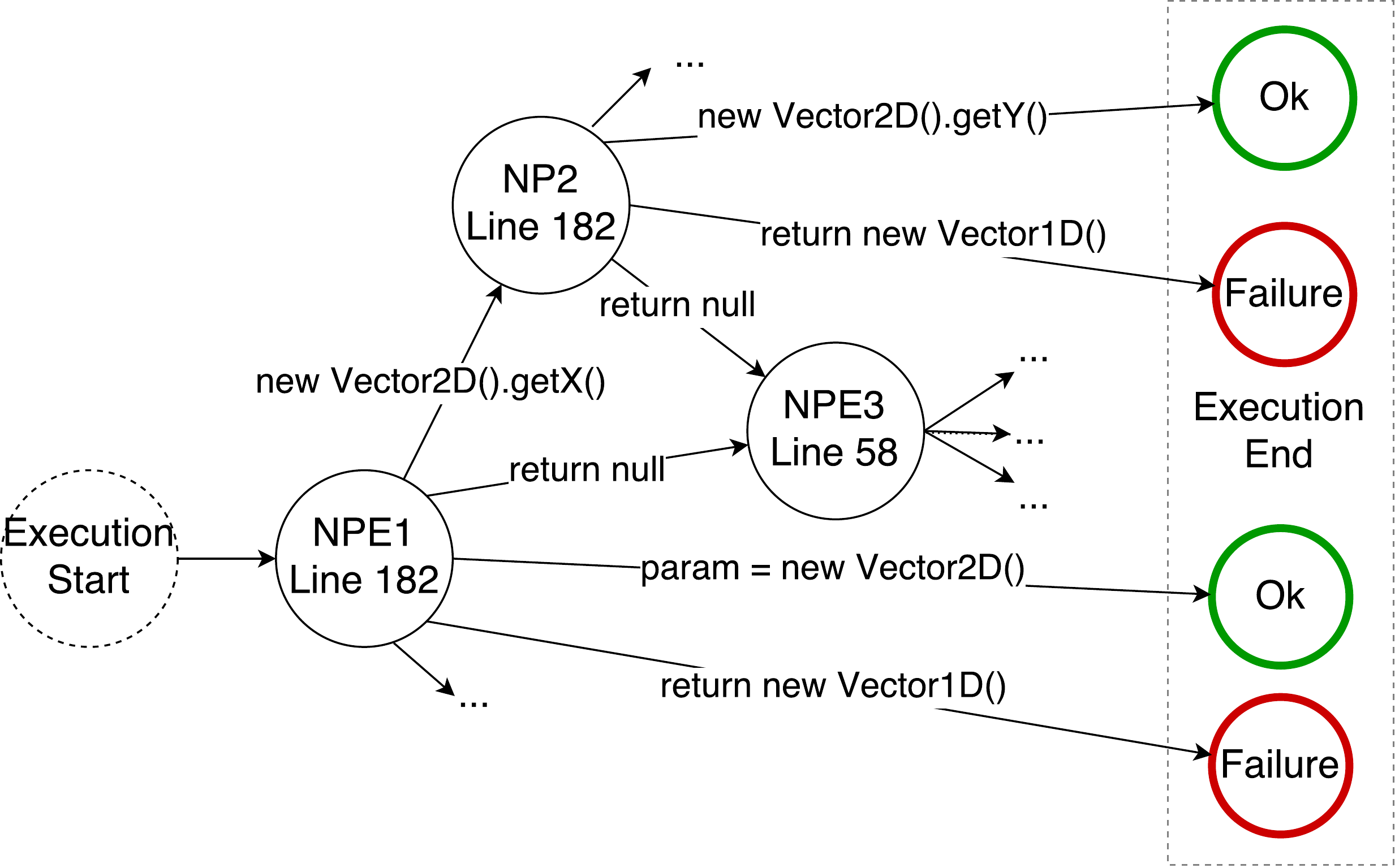}
\caption{Excerpt of the decision tree of Math-988A. One path in this tree is a ``decision sequence'''. 
This figure clearly shows the presence of paths with multiple steps, i.e., the presence of compound decision sequences.}
\label{fig:decisionTree} 
\end{figure}

\Suppressnumber
\begin{lstlisting}[language=diff, caption={The human patch for Math-988A.}, label=lst:Math-988A:human,float]
@@ SubLine.java
(*@\Reactivatenumber{116}@*)
 // compute the intersection on infinite line
 Vector2D v2D = line1.intersection(line2);
+if (v2D == null) {
+   return null;
+}

 // check location of point with respect to first sub-line
 Location loc1 = getRemainingRegion().checkPoint(line1.toSubSpace(v2D));
\end{lstlisting}

\Suppressnumber
\begin{lstlisting}[language=java, caption={The decision points of Math-988A.}, label=lst:Math-988A:buggy,float]
@@ Line.java (*@\Reactivatenumber{181}@*)
Vector1D toSubSpace(Vector2D point) {
  return new Vector1D(cos * point.getX() /* NPE1*/ + sin * p2.getY() /* NPE2*/);
}(*@\Suppressnumber@*)

@@ OrientedPoint.java (*@\Reactivatenumber{57}@*)
double getOffset(Vector2D point) {
  double delta = ((Vector1D) point).getX() /* NPE3*/ - location.getX();
  return direct ? delta : -delta;
}
\end{lstlisting}

\Suppressnumber
\begin{lstlisting}[language=diff, caption={The patch equivalent to a valid failure-oblivious decision taken for Math-988A.}, label=lst:Math-988A:failure-oblivious,float=t]
@@ Line.java(*@\Reactivatenumber{181}@*)
Vector1D toSubSpace(Vector2D point) {
+ if (point == null) {
+    return Vector1D.NaN;
+ }
  return new Vector1D(cos * point.getX()
     + sin * p2.getY());
}
\end{lstlisting}

\textit{Example of Math-988A}:
\label{sec:case-study}
Now we illustrate the disparity of the number of decision sequences for the bug Math-988A.
The initial null pointer exception is triggered in the return statement of method \mycode{toSubSpace} which returns an object of type Vector1D (see line 182 in \autoref{lst:Math-988A:buggy}).
The null pointer exception is triggered when a Vector2D parameter is null and methods \mycode{getX} and \mycode{getY} are called on it. 
The Vector2D parameter is computed by the line 117 in \autoref{lst:Math-988A:human} and then passes as argument to the method \mycode{toSubSpace} at line 123 in \autoref{lst:Math-988A:human}. If two lines do not have an intersection, the geometrical computation of the intersection of them in line 117 of \autoref{lst:Math-988A:human} is null and will then cause the null pointer exception. 

There are three failure points for this bug as shown in \autoref{tab:space}, which means that there are potentially between 1 and 3 null pointer exceptions happening during the execution of the failure input depending on the selected decisions. The source code of these failure points are shown in \autoref{lst:Math-988A:buggy}, and the different decision sequences are illustrated in the decision tree in \autoref{fig:decisionTree}. In this figure, each node represents a failure point, each arrow represents a decision, and each path between the Execution Start and the Execution End represents a decision sequence.

For Math-988A, there are different kinds of failure-oblivious decision:
\begin{enumerate*}
\item initialize the null parameter with a new instance (1 decision point),
\item use a new and disposable instance of Vector2D at both places where the null parameter is used (2 decision points),
\item return null either at the first NPE location or at the second one, triggering another decision in the caller (between 2 and 3 decision points), and
\item return a new instance Vector1D (1 decision point).
\end{enumerate*}

For Math-988A, the reproducing test contains JUnit assertions for checking the expected correct behavior, which should return null when no intersection exists.
In case the failure-oblivious decisions pass those assertions, it means that failure-oblivious computing achieves full correctness and these decisions are illustrated with ``OK'' in \autoref{fig:decisionTree}. 
Otherwise if the failure-oblivious decisions fail on the assertions or trigger another failure, then these decisions do not fully achieve the purpose of failure-oblivious computing and are illustrated with ``Failure'' in \autoref{fig:decisionTree}.
As an example, \autoref{lst:Math-988A:failure-oblivious} shows a valid failure-oblivious decision written in a patch equivalent form.

\answer{3}{For $5/\nbBugs$ failures of our benchmark, the search space contains composite failure-oblivious decision sequences that have more than one decision. For $3/\nbBugs$ failures, the possible failure-oblivious decision sequences have disparate sizes, and our protocol enables us to identify all valid failure-oblivious decision sequences.}

\section{Threats to validity}
\label{sec:threats_validity}

We now discuss the threats to validity of our experiment.
First, let us discuss internal validity. Our experiment is of computational nature, and consequently, a bug in our code may threaten the validity of our results. However, since all our experiment code is publicly available for sake of open-science\footnote{ \url{https://github.com/Spirals-Team/runtime-repair-experiments}}, future researchers will be available to identify these potential bugs.

Second, a threat to the external validity relates to the number of failures considered. Our experiment has considered \nbBugs{} failures from 6 different software projects. Recall that reproducing field failures is a very costly task and consequently, there is a research trade-off between cost and external validity. However, our experiment considers as many or more failures than the related work on failure-oblivious computing.
For external validity, it may also be asked whether our results are specific to Java and whether the search space of failure-oblivious computing search space has the same structure in other programming languages and runtime environments. In our opinion, this is an interesting threat to external validity that calls for more research in this area.

\section{Related work}\label{sec:rw}

Long and Rinard \cite{LongICSE2016} study the search space of patch generation systems. In our work, we consider failure-oblivious decision sequences, which are fundamentally different: while a code patch is a permanent modification to the behavior, a failure-oblivious decision sequence only impacts one single execution, with no effect or regression on subsequent executions, even if they execute the same statement. What’s interesting is that in both cases, contrary to the initial intuition of the research community, there is a multiplicity of possible patches. Long and Rinard’s paper is the first one to study this for static patches, our paper is possibly the first one to comprehensively show that this phenomenon exists for failure-oblivious decision sequences. 

There are several automatic recovery techniques.
One of the earliest techniques is Ammann and Knight's ``data diversity'' \cite{ammann1988data}, that aims at enabling the computation of a program in the presence of failures.
The idea of data diversity is that, when a failure occurs, the input data is changed so that the new input resulting from the change does not result in the failure.
The assumption is that the output based on this artificial input, through an inverse transformation, remains acceptable in the domain under consideration.
The input transformations can be seen as a kind of failure-oblivious model. As such, our protocol could be used to reason on the search space of data diversity.

Demsky et al. \cite{demsky2003automatic} present a language for the specification of data structure invariants. The invariant specification is used to verify and repair the consistency of data structure instances at runtime. 
In their work, Demsky et al. do not study the associated search space.

Rinard et al. \cite{rinard2004enhancing} presents a technique to avoid illegal memory accesses by adding additional code around each memory operation during the compilation process.
For example, the additional code verifies at runtime that the program only uses the allocated memory.
If the memory access is outside the allocated memory, the access is ignored instead crashing with a segmentation fault.
We apply different decisions to handle a given failure (and not a single code, hard-coded in the injected code), and we use an oracle to reason about the viability of the decision.

Perkins et al. \cite{perkins2009automatically} proposes ClearView, a system for automatically handling errors in production.
The system consists of monitoring the system execution on low-level registers to learn invariants. 
Those invariants are then monitored, and if a violation of an invariant is detected ClearView forces the restoration.
From an engineering perspective, the difference is we reason on decision sequences, while ClearView analyzes each decision in isolation.
From a scientific perspective, our work finely characterizes the search space and the outcomes of failure-oblivious computing based on execution modification. 

Rx \cite{qin2005rx} is a runtime repair system based on changing the environment upon failures. Rx employs checkpoint-and-rollback for re-executing the buggy code when failures happen. The differences are as follows: 1) Rx does not change the execution itself but the environment 2) the search space of Rx is smaller (a set of predefined strategies) 3) Rx’ experiment does not include systematic exploration of the search space.

Kling et al. \cite{kling2012bolt} propose Bold a system to detect and escape infinite and long-running loops.
On user demand, Bolt is attached to a running application and tries different strategies to escape the infinite loop. 
If a strategy fails, Bolt uses rollback to restore the state of the application and then tries the next strategy.
Bolt does not reason about decision sequences as we do in this paper.

Long et al. \cite{long2014automatic} introduces the idea of “recovery shepherding” in a system called RCV.
Upon certain errors (null dereferences and divide by zero), recovery shepherding consists in returning a manufactured value, as for failure-oblivious computing.
The key idea of recovery shepherding is to track the manufactured values so as to see 1) whether they are passed to system calls or files and 2) whether they disappear.
The key difference with our work lies in the reasoning about the effect of the combinations (by storing and keeping information about the actual valid decision sequences). 

Jula et al. \cite{jula2008deadlock} presents a system to defend against deadlocks at runtime.
The system first detects synchronization patterns of deadlocks, and when the pattern is detected, the system avoids re-occurrences of the deadlock with additional locks.
The pattern detection is related to the detector of instances of the fault model under consideration. However, Jula et al. do not explore and compare alternative locking strategies. We note that our protocol may be plugged on top of their systems to explore the search space of locking sequences. 

Hosek and Cadar \cite{hosek2013safe} switch between application versions when a bug is detected.
This technique can handle failures because some bugs disappear while others appear between versions. 
We can also use our protocol to systematically explore the sequences of runtime jumps across versions. 

Assure \cite{sidiroglou2009assure} is a self-healing system based on checkpointing and error virtualization.
Error virtualization consists of handling an unknown and unrecoverable error with error handling code that is already present in the system yet designed for handling other errors. 
While Assure does self-healing by opportunistic reuse of already present recovery code, our failure-oblivious model handles failures by modifying the state or flow.

Carzaniga et al. \cite{carzaniga2010automatic} repair web applications at runtime with a set of manually written, API-specific alternatives rules. This set can be seen as a hardcoded set of failure-oblivious decision sequences.
On the contrary, we do not require a list of alternatives but instead relies on an abstract failure-oblivious model that is automatically instantiated at runtime.

Berger and Zorn \cite{berger2006diehard} show that is possible to effectively tolerate
memory errors and provide probabilistic memory safety by randomizing the memory allocation and providing memory replication.
Exterminator \cite{novark2007exterminator} provides more sophisticated fault tolerance than \cite{berger2006diehard} by performing fault localization before applying memory padding.
The work by Qin et al. \cite{qin2005safemem} exploits a specific hardware feature called ECC-memory for detecting illegal memory accesses at runtime. The idea of the paper is to use the consistency checks of the ECC-memory to detect illegal memory accesses (for instance due to buffer overflow). 
Both techniques are semantically equivalent in the normal case. We have reasoned about the search space of execution modifications that are not semantically equivalent, where one taken decision can impact the rest of the computation. 

Dobolyi and Weimer~\cite{dobolyi2008changing} present a technique to tolerate null dereferences.
Using code transformation, they introduce hooks to a recovery framework.
This framework is responsible for forwarding recovery of the form of creating a default object of an appropriate type of skipping instructions.
Kent~\cite{kent2008dynamic} proposes alternatives to null pointer exceptions.
He proposes to skip the failure line or exits the method by a return when a null pointer exception is detected.
In those two contributions, there is no reasoning on the search space of failure-oblivious computing, as done in this work.

Jeffrey et al. \cite{jeffrey2010execution} present a technique to assist the developers to locate the root cause of memory errors.
In this work Jeffrey at al. suppress the execution of the statement that produces the failure and repeats this procedure until the execution of the program does not fail. 
The last suppressed statement should according to Jeffrey at al. be close to the root cause of the memory error.
This approach uses the failure-oblivious strategy to continue the execution of the program in order to gain knowledge. In this case, they want to identify the root cause of the memory error.
The main difference is that Jeffrey et al. don't use the failure-oblivious technique to fix the application but to get knowledge during the execution of the program.
We focus on the failure-oblivious computing search space to understand the failure-oblivious behavior and we also consider different failure-oblivious strategies to handle failures.

\section{Conclusion}
\label{sec:conclusion}

In this paper, we characterized the key concept of search space of failure-oblivious computing.
In order to understand the behavior of failure-oblivious computing, we proposed an algorithm to exhaustively explore this search space.
We performed an empirical study on \nbBugs real java bugs to draw a precise picture of the nature of the failure-oblivious search-space on reals bugs.
We find out that there are several possible execution modifications to handle a failure and several execution modifications have to be taken in a row to handle some specific failures.

We have now a better understating of the size of the failure-oblivious search space, we plan to develop techniques to select better state modifications in the future so that the capability of failure-oblivious system in production environment can be improved. 

\balance
\bibliographystyle{IEEEtran}
\bibliography{references}

\end{document}